\documentclass[submission,copyright,creativecommons]{eptcs}
\providecommand{\event}{BEAT 2014} 
\usepackage{amssymb}
\usepackage{bm}
\usepackage{graphicx}
\usepackage{url}
\usepackage{color}
\usepackage{comment}
\usepackage{times}
\usepackage{amsmath}
\usepackage{esvect}
\usepackage{listings}
\usepackage{subfigure}
\usepackage{space}

\newcommand{\laucom}[1]{}

\newcommand{\participant}[1]{\ensuremath{\mathtt{#1}}}

\newcommand{\s}{\ensuremath{s}}

\newcommand{\mqueue}[2]{\ensuremath{#1 : #2}}
\newcommand{\emptyqueue}[1]{\mqueue{\s}{\emptyset}}

\newif\ifmr
\mrfalse 

\newcommand{\ptp}[1]{{\participant{#1}}}

\newcommand{\node}{\mathtt{n}}

\newcommand{\dc}[1]{\delta_{#1}}

\newcommand{\cvar}{x}
\newcommand{\cof}[1]{\cvar_{#1}}

\newcommand{\CODE}[1]{\texttt{\CODESIZE#1}}
\newcommand{\CODESIZE}{\fontsize{8}{4}\selectfont}  
\newcommand{\CODESTYLE}{\ttfamily}

\definecolor{dkblue}{rgb}{0,0.1,0.5}
\definecolor{dkgreen}{rgb}{0,0.4,0}
\definecolor{dkred}{rgb}{0.4,0,0}

\lstnewenvironment{PYTHONLISTING}%
{
\lstset{
  language=python,
  showstringspaces=false,
  formfeed=\newpage,
  tabsize=2,
  commentstyle=\color{dkgreen},
  basicstyle=\CODESTYLE\fontsize{6}{6}\selectfont,
  keywordstyle=\color{dkblue},
  emphstyle=\color{dkblue}\bfseries,
  morekeywords={models, lambda, forms, def, class, t},
  morekeywords=[2]{int,string},
  emph={access,and,as,break,class,continue,def,del,elif,else,%
	except,exec,finally,for,from,global,if,import,in,is,%
	lambda,not,or,print,raise,return,try,while,assert,with, from, at, to, rec, choice, protocol,role},
  morecomment=[s]{[}{]},
  escapeinside={(*@}{@*)}
}
}
{
}

\title{Timed Runtime Monitoring for Multiparty Conversations}

\author{Rumyana Neykova
\institute{Imperial College London, UK}
\and
Laura Bocchi 
\institute{Imperial College London, UK}
\and
Nobuko Yoshida
\institute{Imperial College London, UK}
}

\def\titlerunning{Timed Runtime Monitoring for Multiparty Conversations}
\def\authorrunning{R. Neykova, L. Bocchi \& N. Yoshida}
\begin{document}
\maketitle

\begin{abstract}
We propose a dynamic verification framework for protocols in real-time distributed systems. 
The framework is based on Scribble, a tool-chain for design and verification of choreographies based on multiparty session types,  
developed with our industrial partners. Drawing from recent work on multiparty session types for real-time interactions, 
we extend Scribble with clocks, resets, and clock predicates constraining the times in which interactions should occur. We present a timed API for Python to program distributed implementations of Scribble specifications. A dynamic verification framework ensures the safe execution of applications written with our timed API: we have implemented dedicated runtime monitors that check that each interaction occurs at a correct timing with respect to the corresponding Scribble specification. 
The performance of our implementation and its practicability are analysed via benchmarking. 
\end{abstract}

\label{sec_introduction}
Recent work \cite{BYY14} extends Multiparty Session Types (MPSTs) to enable the verification of real-time distributed systems.  
This timed extension allows to express properties on the causalities of interactions, on the carried datatypes, and on the \emph{times} in which interactions occur. 
In this paper, we apply the theory in \cite{BYY14} to implement a toolchain for specification and runtime verification of real-time interactions, and evaluate our prototype implementation via benchmarking.  

This work is motivated by our collaboration with the Ocean Observatories Initiative (OOI) \cite{OOI}, directed at developing a large-scale cyber-infrastructure for ocean observation. The type of protocol used in the governance of the OOI infrastructure (e.g., users remotely accessing instruments via service agents) can be suitably expressed using MPSTs, and an \emph{untimed} monitoring framework based on MPSTs \cite{HuNYDH13} is now integrated into OOI. 
%
Time, however, in necessary in many OOI use-cases, for instance to associate timeouts to requests when resources can be used for fixed amounts of time, or to schedule the execution of services at certain time intervals to reduce the busy wait and minimise energy consumption). 



\section{Running example and methodology}
\label{sec_background}

\begin{figure}
\begin{center}
\includegraphics[scale=0.45]{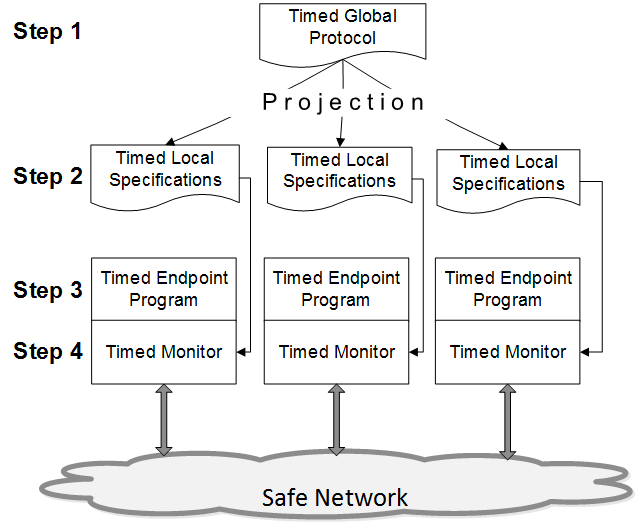}\hspace{0.5cm}
\includegraphics[scale=0.38]{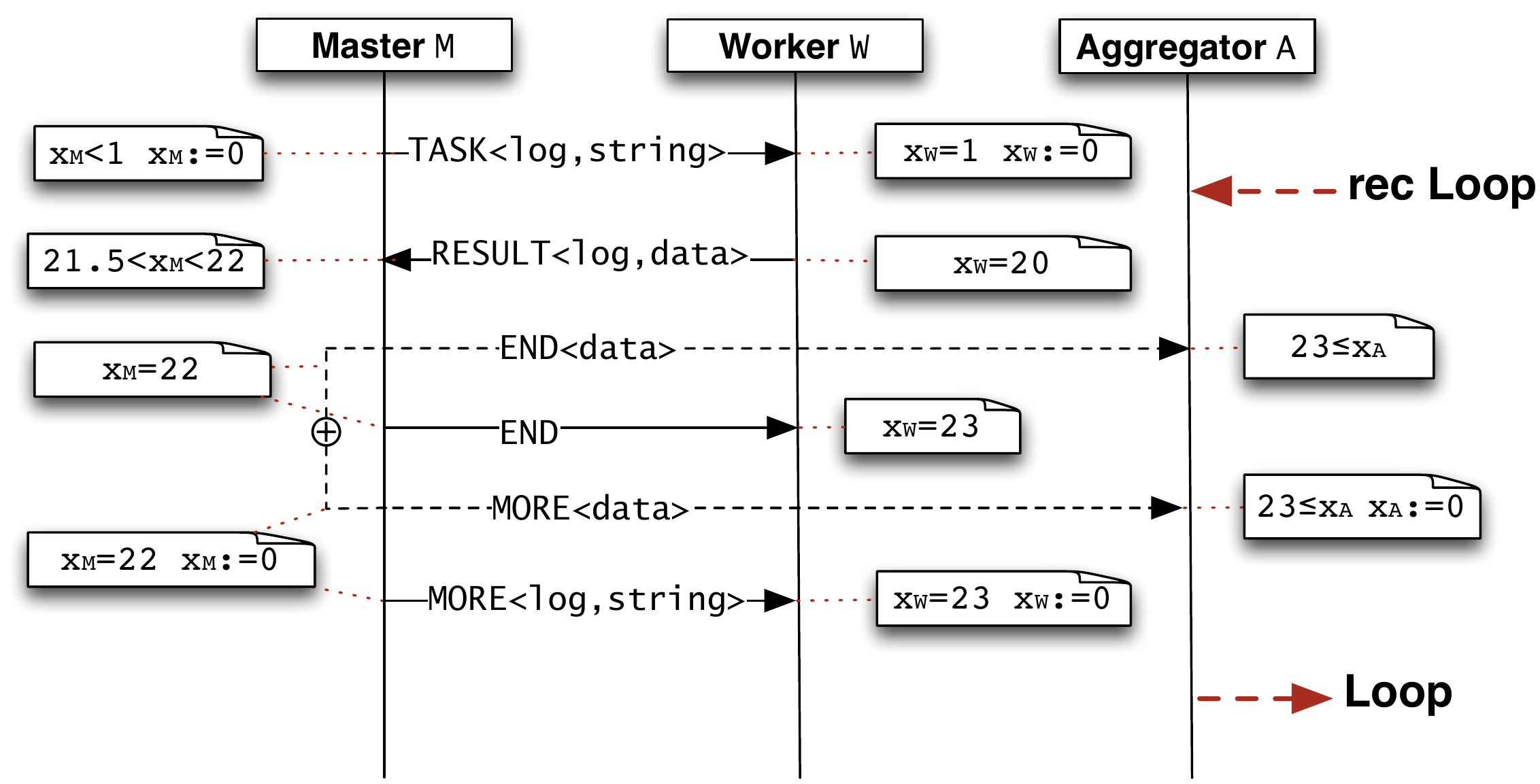}
\caption{Scribble toolchain framework (left) and global protocol for log crawling (right)}
\label{fig_mpst}
\end{center}
\end{figure}

Our toolchain centres on a specification language called Scribble~\cite{scribble,scribble10}, and 
supports the top-down development methodology illustrated in Figure~\ref{fig_mpst} (left). 
In \textbf{step 1}, a global communication is \emph{specified} as a Scribble \emph{timed global protocol}. A timed global protocols defines: 
(a) the causality among interactions in a session involving two or more participants, (b) the datatypes carried by the messages, and (c) the timing constraints of each interaction. We extended Scribble with the notion of time from~\cite{CTA,BYY14}: each participant owns a clock on which timing constraints can be defined. The clock can be reset many times in a session, and we assume that time flows at the same pace for all participants. 
%
In \textbf{step 2}, the Scribble toolchain is used to algorithmically \emph{project} the timed global protocol to \emph{timed local protocols}. Each timed local protocol specifies the actions in a session (and their timing) from the perspective of a single participant. 
In \textbf{step 3}, principals over a network \emph{implement} one or more, possibly interleaved, timed local protocols. We will call these implementations \emph{timed endpoint programs}. In our prototype implementation, timed local protocols are written in native Python using our in-house developed Conversation API. Our Python conversation API is a message passing library that supports the core primitives for communication programming of MPSTs. 
Finally, in \textbf{step 4}, the timed endpoint programs are executed. Each endpoint is associated to a dedicated and trusted monitor. A monitor checks that the interactions of the monitored timed endpoint program conform to the implemented timed local protocols. In case of violation, the monitor either throws a time error (error detection mode), or triggers recovery actions to amend the conversation (error prevention/recovery mode).

%



\paragraph{Outline of the paper.} In the following sections we will discuss in
detail each of the steps of the methodology illustrated in
Figure~\ref{fig_mpst} (left). 
In \S~\ref{sec_specification} (\textbf{steps 1 and 2}) we present 
Scribble timed global and local protocols, which are a practical and more human-readable incarnation of timed global and local types in~\cite{BYY14}. 
We implemented the algorithms in \cite{TRTime} to check two time-consistency properties over timed global types/protocols: feasibility and well-formedness~\cite{BYY14}. The projection of Scribble timed global protocols  has also been implemented following~\cite{BYY14,TRTime}.
In  \S~\ref{sec_implementation} we present our timed API (\textbf{step 3})
 based on the calculus with delays in~\cite{BYY14} (a simple timed
 extension of the $\pi$-calculus used to implement timed local
 types). In \S~\ref{sec_enforcement} we discuss runtime enforcement of
 timed properties (\textbf{step 4}). 
Timed local protocols are automatically encoded into timed automata
(using the encoding from timed local types to timed automata presented
in~\cite{BYY14,TRTime}), which are in turn used by our runtime
monitors for error detection. Additional mechanisms for error
prevention and recovery are implemented and explained. Benchmarks are in \S~\ref{sec_benchmark} and related work in~\S~\ref{sec_conclusion}. Our prototype implementation is available at \cite{py_time}.


\section{Specifying Timed Protocols with Scribble}
\label{sec_specification}
\paragraph{Running example.} We present our framework via a running example: a protocol for distributed computation of a word count over a set of logs. The timed global protocol, informally illustrated in Figure ~\ref{fig_mpst} (right), involves three participants: a master $\ptp M$, a worker $\ptp W$ and an aggregator $\ptp A$. Each participant has a clock, $\cof{\ptp M}$, $\cof{\ptp W}$, and $\cof{\ptp A}$, respectively, initially set to $0$.\footnote{As customary in MPSTs, protocols start synchronously for all participants, hence all clocks start counting at the same time.}
At the beginning of the session $\ptp M$  sends $\ptp W$  a message of type $\mathsf{TASK}$ together with a variable of type $\mathtt{log}$ (i.e., the list of log names to crawl) and a variable of type $\mathtt{string}$ (i.e., the word to search). 
The message must be sent by $\ptp M$ within one second ($\cof{\ptp M}<1$) and received by $\ptp W$ at time $\cof{\ptp W}=1$. Both $\ptp M$ and $\ptp W$ reset their clocks upon sending/receiving the message.
The protocol then enters a loop. At each iteration,  $\ptp W$ replies to $\ptp M$ in exactly 20 seconds with a message of type $\mathsf{RESULT}$ along with a variable of type $\mathtt{log}$ (i.e., the logs that have been crawled in the given amount of time) and a variable of type $\mathtt{data}$ (i.e., the result of the word search). This message is received by $\ptp M$ at any time satisfying $21.5<\cof{\ptp M}<22$.
A choice is then made locally to $\ptp M$ at time $22$: depending on whether the results are satisfactory or not, the worker chooses to either 
terminate the session (message of type $\mathsf{END}$), or to continue the crawling (messages of type $\mathsf{MORE}$). If $\ptp{W}$ chooses $\mathsf{MORE}$ all clocks are reset. In both cases the results of the last iteration are forwarded to $\ptp A$. 
This timed protocol allows ${\ptp M}$ to wake up at regular intervals (e.g., every $20$ seconds) to evaluate the results and decide when to continue or terminate the loop. Otherwise $\ptp M$ can remain idle (e.g., sleep).

%

\paragraph{Timed global protocols} 
We let each participant (or {role}) in a Scribble timed global protocol to \emph{own} one real-valued clock which can be reset many times. The (asynchronous) interactions between pairs of participants can be thought as being broken down into two actions: the sending action and the receiving action. Each sending (resp. receiving) action is annotated with a constraint and a reset predicate, both defined on the clock owned by the sender (resp. receiver). An action can be executed only if the associated constraint is satisfied, and after its execution the clock of the sender/receiver is reset according to the reset predicate. 
Clock constraints and reset predicates are represented in Scribble as annotations on the message interactions, 
enclosed by square brackets and are explicitly bound to a participant. Figure~\ref{scribble_global} (left) shows the Scribble timed global protocol of the example in Figure~\ref{fig_mpst} (right). 
 
\begin{figure}
\begin{minipage}[t]{0.55\linewidth}
\begin{PYTHONLISTING}
global protocol WordCount at M(role A, role W)
    [xm@M: xm<1,reset(xm)][xw@W: xw=1,reset(xw)]
    task(log,string) from M to W;
    rec Loop{
        [xw@W: xw=20][xm@M: 21.5<xm<22]
        result(data) from W to M;
        choice at M{
        	[x@M: xm=22][xa@A: 23<=xa,reset(xa)]
        	more(data) from M to A;
        	[xm@M: xm=22,reset(xm)][xw@W: xw=23,reset(xw)]
        	more(log,string) from M to W;    
        	continue Loop;
        } or {
	        [xm@M: xm=22][xa@A: 23<=xa]        
        	end(data) from M to A;
        	[xm@M: xm=22][xw@W: xw=23]
        	end() from M to W; } }
\end{PYTHONLISTING}
\end{minipage}
\begin{minipage}[t]{0.44\linewidth}
\begin{PYTHONLISTING}
local protocol WordCount at M(role A, role W)
    [xm@M: xm<1,reset(xm)]
    task(log,string) to W;
    rec Loop{
        [xm@M: 21.5<xm<22]
        result(data) from W;
        choice at M{
        	[xm@M: xm=22]
        	more(data) to A;
        	[xm@M: xm=22, reset(xm)]
        	more(log,string) to W;    
        	continue Loop;
        } or {
	        [xm@M: xm=22]        
        	end(data) to A;
        	[mx@M: xm=22, reset(xm)]
        	end() to W; } }
\end{PYTHONLISTING}
\end{minipage}
\caption{Scribble timed global (left) and local (right) protocol for $\ptp M$}
\label{scribble_global}
\end{figure}




\paragraph{Timed properties of global protocols} The theoretical framework in~\cite{BYY14} sets two time-consistency conditions on timed global types: \emph{feasibility} (first introduced in~\cite{Apt:1987:AFD:41625.41642}) requiring that for each partial execution allowed by a specification there is a correct complete one, and \emph{wait-freedom} requiring that if senders respect their time constraints, then receivers never have to wait for their messages. These conditions rule out protocols which may intrinsically lead to violations, as shown by the examples below. 


\begin{minipage}{0.52\linewidth}
\begin{PYTHONLISTING}
global protocol fooBar (role A, role B)
    [xa@A: xa<10][xb@B: xb<5]
    msg(string) from A to B;
    ...
\end{PYTHONLISTING} 
\end{minipage}
\begin{minipage}{0.4\linewidth}
\begin{PYTHONLISTING}
global protocol fooBar (role A, role B)
    [xa@A: x<10][xb@B: x<20]
    M1(string) from A to B;
    [xb@B: x<20][xa@A: true]
    M2(string) from B to A;
    ...
\end{PYTHONLISTING}
\end{minipage}

The protocol on the left violates feasibility since it allows $\ptp A$ to send $\mathtt{msg}$ at any time satisfying $\mathtt{xa}<10$, for instance at time $8$, for which then $\ptp B$ has no means to satisfy constraint $\mathtt{xb}<5$ for the corresponding receive action. 
The protocol on the right violates wait-freedom. Assume $\ptp B$ to be implemented by a timed endpoint program that receives $\mathsf{M1}$ at time $5$, and then engages in a time-consuming activity for $14$ seconds before sending $\mathsf{M2}$. The plan of $\ptp B$ conforms to the corresponding timed local protocol. If, however, we compose the timed endpoint program described before with an implementation of $\ptp A$ that sends $\mathsf{M1}$ at time $8$, we have that $\ptp B$ will not find the message in the queue at the expected time $5$, will `get late' with respect to his planned timing, and may end up violating the contract at a later action.   

We implemented a syntactic checker of {feasibility} and {wait-freedom}, based on the algorithms given in~\cite{TRTime}. 
The algorithms are based on a directed acyclic graph where: (i) nodes model the actions of (the one-time unfolding of) a timed global protocol and are annotated with the clock constraints and reset predicates of that action, and (ii) edges model the temporal/causal dependencies between actions. For each node $\node$ we build a \emph{dependency constraint} $\dc{\node}$, using the information on constraints and resets in the path to $\node$, to model the range of `absolute' times in which the the state represented by $\node$ can be reached. For feasibility we check that the clock constraint $\delta$ annotating each node $\node$ admits some solution on or after any time allowed by $\dc{\node}$; for wait-freedom, we check that all the solutions of $\delta$ occur at the same or at a later time with respect to any time allowed by $\dc{\node}$.

In~\cite{BYY14} these conditions yield progress for \emph{statically} validated programs. This is not the case for \emph{dynamically} verified programs against MPSTs~\cite{FORTE13} since monitors do not enforce interactions on participants that deliberately refuse or cannot (e.g., their machine is down) send the remaining messages in a protocol. Ensuring that conversations are established on feasible and wait-free protocols is, however, a good practice as it prevents progress violations are induced by the protocol itself.

\paragraph{Timed local protocols.} After being checked for feasibility and wait-freedom, the timed global protocol is automatically projected to timed local protocols, one for each participant. Figure ~\ref{scribble_global} (right) presents the projection (a timed local protocol) into $\ptp M$. A timed local protocol is essentially a view of the timed global protocol from the perspective of one participant. By decomposing the timed global protocol into separate but consistent timed local protocols, projection is a key mechanism to enable distributed enforcement of global properties in our framework. 

\section{Implementing Timed Protocols with Python}
\label{sec_implementation}
When implementing a Scribble timed local protocol -- \textbf{step 3} in Figure~\ref{fig_mpst} (right) -- one must take care that actions will be executed at the right times. 
We present a timed conversation API for real-time processes in Python which allows programmers to 
(1) delay the execution of an action to match a prescribed timing while avoiding busy wait, and (2) interrupt an ongoing computation to meet an approaching deadline.




\paragraph{Idle delays.}
In~\cite{BYY14} processes are modelled using a simple extension of the $\pi$-calculus with a delay operator $\mathtt{delay}(\mathtt{t}).P$ that executes as process $P$ after waiting exactly $\mathtt{t}$ units of time. All the other actions are assumed to take no time. 
Our API is designed following a similar approach: we introduce a primitive for time-passing and assume that all other operations take no time. Whereas, in practice, Python operations always take some time, we assume that these delays are negligible w.r.t.  those explicitly modelled in the constraints of a Scribble timed protocol. 
We consider non negligible delays expressed in seconds, whereas the other Python and communication operations usually take times in the order of milliseconds. In the runtime verification of a Python process we let the monitor neglect time discrepancies in the order of milliseconds. For example, when running our prototype, the monitor considers satisfactory a scenario where a clock $x$ has value $18.35001073$ and the constraints requires $x=18$. In the use-cases we considered so far these discrepancies do not create problems since their accumulation in long (e.g., recursive) executions is limited by a careful use of resets. 
Idle delays are expressed using two constructs: \CODE{delay} (relative delay) and \CODE{delay\_until} (absolute delay). The primitives have been implemented using a lower level function provided by the `gevent' library: \CODE{gevent.sleep} which lets time elapse for a specified about of time and (in the case of \CODE{delay\_until}) \CODE{gevent.timeout}.

\paragraph{Computation-intensive functions and timeouts.}
A delay in the calculus in~\cite{TRTime} ($\mathtt{delay}(\mathtt{t}).P$) does not model  only idle waiting, but also busy time spent doing some computation. 
Hereafter we will call \emph{computation-intensive functions} those operation that take an amount of time which is not negligible such as, for instance, the log crawling performed by the worker in our running example. It may be difficult to foresee the exact duration of a computation-intensive function; in order to ensure that its execution does not exceed the time prescribed by the local protocol, we associate each computation-intensive function to a parameter \CODE{timeout} that is an upper bound to the duration of its execution; an exception is raised if the function is not completed in the given time frame. In the running example we use, in the implementation of the worker, the function \CODE{self.crawl(log, word, timeout=20)} which interrupts the crawling after $20$ seconds; the resulting exception can be handled by simply proceeding with the computation while considering the result of the function as `partial'. 

\paragraph{Timed API.}
We illustrate more concretely the primitives introduced earlier in this section through a Python implementation of the running example.  Figure~\ref{master_role_python} shows the Python program for the participants of our running example. 
\begin{figure}	
\vspace{-5mm}
\begin{minipage}[t]{0.33\linewidth}
\begin{PYTHONLISTING}
def master_proc():
    c = Conversation.create(...)
    c.send('W', 'task', 'log', 'string')
    c.delay(22)
    c.receive('W')
    while more_tasks():	
    	c.send('A','more', 'data')
    	c.send('W', 'more', 'log', 
    		   'string')
    	c.delay(22)
    	c.receive('W')
    c.send('A', 'end', 'data')
    c.send('W', 'end')
\end{PYTHONLISTING}
\end{minipage}
\begin{minipage}[t]{0.33\linewidth}
\begin{PYTHONLISTING}
def worker_proc():
    c = Conversation.join(...)
    c.delay(1)
    log = c.receive('M')
    while conv_msg.label != 'end':
        data = self.crawl(log, 
        				  timeout=20)        
        c.send('M','result', data)
        c.delay(23)
        conv_msg = c.receive('M')
\end{PYTHONLISTING}
\end{minipage}
\begin{minipage}[t]{0.33\linewidth}
\begin{PYTHONLISTING}
def aggr_proc():
    c = Conversation.join(...)
    op = None
    while op !='end':
            c.delay(23)
            conv_msg = c.receive('M')
            op = conv_msg.label

\end{PYTHONLISTING}
\end{minipage}
\vspace{-0.4cm}\label{master_role_python}
\caption{Participants implementation in Python}
\end{figure}
The implementation for the master process is given in Figure~\ref{master_role_python} (left). 
Line 1-2 start the conversation on channel \CODE{c}. Then, following the local protocol, the master sends a
request to the worker passing the log name and the word to be counted. The send method, called on conversation channel \textit{c}, takes as arguments the destination role, message operator and payload values. 
This information is encapsulated in the message payload as part of a conversation header and is later used for checking by the runtime verification module. 
The receive method can take the sender as a single argument, or additionally the operator of the desired message. The code continues with the \textit{delay} operator. Meanwhile, other green threads run, preventing the worker from busy waiting. 
The implementation for the worker process is given in Figure~\ref{master_role_python} (centre); in Line 6 
operation \CODE{self.crawl(log, word, timeout=20)} models a computation-intensive function.
The aggregator process is shown in Figure~\ref{master_role_python} (right).



\section{Runtime Verification and Enforcement of Time Properties}
\label{sec_enforcement}
We applied the encoding of MPSTs into Communicating Timed Automata from~\cite{BYY14} to derive runtime monitors for the timed setting. In this section we introduce our monitoring framework and discuss the challenges of monitoring the timing of actions. 
A monitor acts as a membrane between one endpoint and the rest of the network, checking that the send and receive actions performed by that timed endpoint program conform to the implemented Scribble timed local protocols. 
In a network where all endpoints are monitored then either all actions will occur at the prescribed timing, or an error will be detected. 
The monitor has two purposes (or \emph{modes}) w.r.t time: error detection and error prevention/recovery. 


\paragraph{Error detection.} 
The monitor verifies the communication actions of the monitored endpoint against Scribble timed local protocols, expressed as timed automata. 
First, the monitor verifies that the type (operation and payload) of each message matches its specification and that occurs in the right causal order w.r.t. the Scribble protocol (as in the untimed Scribble toolchain). Second, the monitor checks the correct timing of actions. For each ongoing protocol, the monitor is augmented with a \emph{local clock}. 
A synchronisation has been introduced in the prototype to ensure that all processes and monitors will start a protocol at the same time, with clocks set to $0$. When a timed endpoint program executes an action the monitor checks the clock constraint of that action (in the timed automaton) against the value of the local clock. If the action complies with the prescribed timing is made visible (i.e., forwarded) into the network, otherwise the monitor raises a \textit{TimeException}. 
For example, if we change the {delay} of the program in Figure~\ref{master_role_python} (left) to be \CODE{delay(30)} this will result in a \textit{TimeException}. 
Error detection allows rigorous blame-assignment analysis in case of violation (we assume trusted monitoring framework).


\paragraph{Error prevention/recovery.} This mode relies on the error detection mechanism: when a violation occurs the monitor enforces the clock constraints by generating recoverable actions. We have two types of scenarios: an action is launched by the local endpoint too early or too late (or not at all) w.r.t. the prescribed timing.  
In the first case, the monitor generates a \textit{delay} equal to the time that is left until an appropriate time is reached, and then it forwards the action to the rest of the network. For example, if we delete the line  \CODE{delay(20)} in Figure~\ref{master_role_python} (left) or modify it with a smaller delay then the monitor will introduce the missing delay so that the monitored application will appear correct to the network. 
When a deadline is reached but its associated action is still not executed, the monitor raises a \textit{TimeoutException}. The application can try and recover itself using the exception handler, e.g., by interrupting an ongoing computation and continuing the conversation, or restarting the protocol with different settings. 

The monitor looks at the next action prescribed by the timed automaton (or \emph{prescribed action})  and acts according to the pre- and post-actions in the table. Pre-actions (resp. post-actions) denote actions performed by the monitor before (resp. after) that the timed endpoint program executes the action that corresponds to the prescribed action. The table below 
summarises the actions generated by the monitor in error prevention/recovery mode. 
 \begin{center}
 \footnotesize
\begin{tabular}{ l | l  | l | l }
prescribed action & clock constraint & pre-action & post-action \\ \hline 
\CODE{s.send} & \CODE{$x \geq n$} &               & \CODE{s.sleep($n-x_{cur})$}\\
\CODE{s.send} & \CODE{$x \leq n$} & \CODE{s.timeout($n-x_{cur}$)}      & \\
\CODE{s.recv} & \CODE{$x \geq n$} & \CODE{s.sleep($n-x_{cur}$)} & \\
\CODE{s.recv} & \CODE{$x \leq n$} &               & \CODE{s.timeout($n-x_{cur}$)} 
\end{tabular}
\end{center}
In the table $x_{\textit{cur}}$ is the local clock of the monitor. 
If the clock constraint of the prescribed action specifies a lower bound $x\geq n$ then the monitor introduces a delay of exactly $n$ (mapped to the low level Python \textit{gevent.sleep} primitive). In case of send we have a post-action: the monitor sleeps \emph{after} observing the action of the endpoint  and forwards it to the network at the right time. In case of receive we have a pre-action: the monitor sleeps \emph{before} observing the receive action so that the incoming message will be read at the appropriate time.
Similarly, when the clock constraint specifies an upper bound $x\leq n$ the monitor inserts a \textit{timeout} (a timer triggering a \textit{TimeoutException}). 

\section{Benchmarks on Transparency of Timed Monitors}
\label{sec_benchmark}
 \begin{figure}[t]
 \vspace{-10mm}
\begin{minipage}{0.48\linewidth}
\begin{PYTHONLISTING}
global protocol WordCount at M(
	role R, role W)
    [xm@M: xm<0.01,reset(xm)][xw@W: xw=0.01,reset(xw)]
    task(log,tring) from M to W;
    rec Loop{
        [xw@W: xw=0.20][xm@M: 0.21<xm<0.22]
        result(data) from W to M;
        choice at M{
        	[x@M: xm=0.22][xa@A: 0.23<=xa,reset(xa)]
        	more(data) from M to A;
        	[xm@M: xm=0.22,reset(xm)]
        	[xw@W: xw=0.23,reset(xw)]
        	more(log,string) from M to W;    
        	continue Loop;
        } or {s
	        [xm@M: xm=0.22][xa@A: 0.23<=xa]        
        	end(data) from M to A;
        	[xm@M: xm=0.22][xw@W: xw=0.23]
        	end() from M to W; } }
\end{PYTHONLISTING}
\end{minipage}
\begin{minipage}{0.5\linewidth}
\includegraphics[scale=0.27]{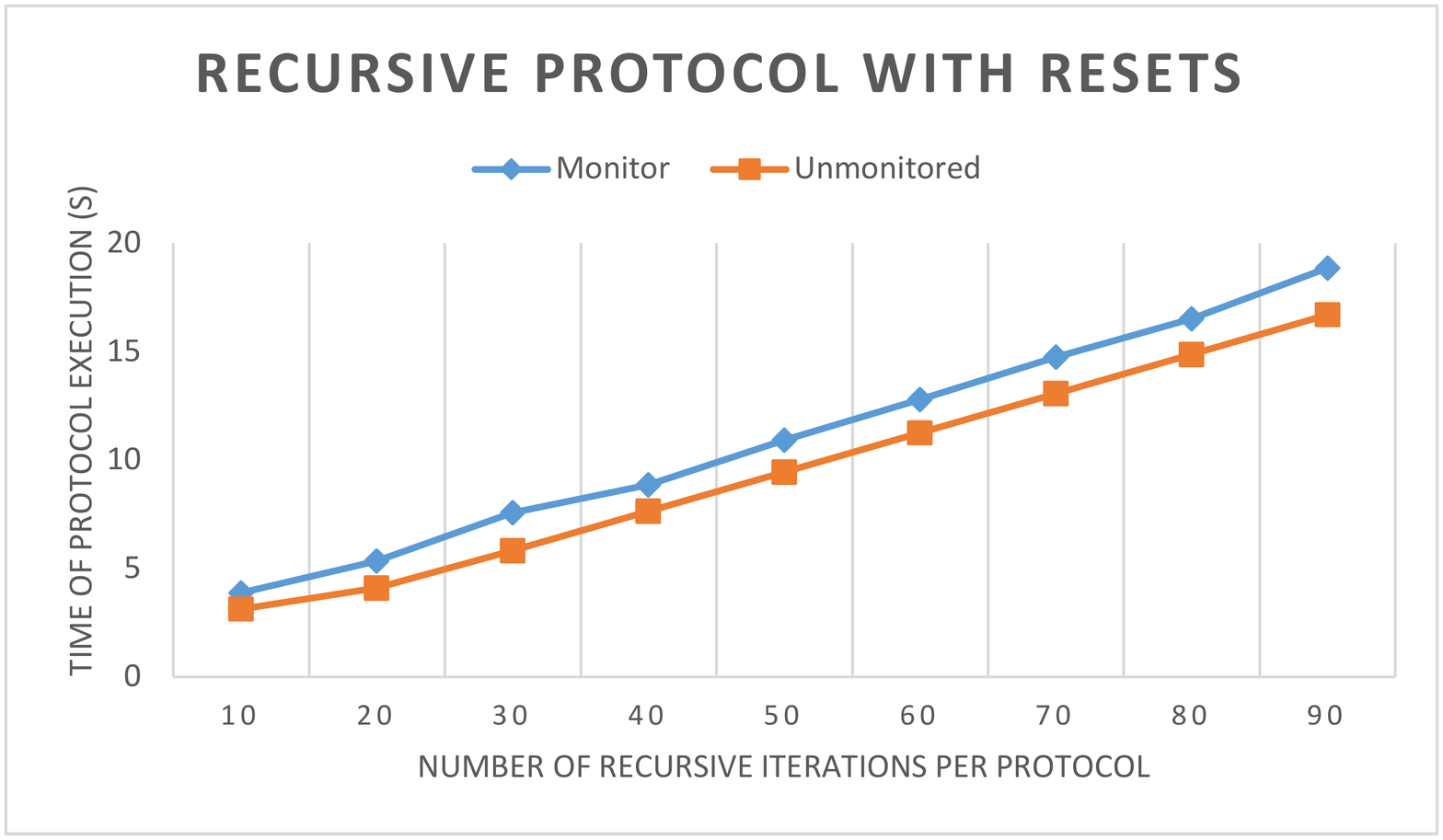}
\end{minipage}
\label{reset}
\vspace{-8mm}
\caption{Recursive protocol with resets (left) and its execution time per number of recursions (right)}

\end{figure}
The practicality of our timed monitoring framework depends on the transparency of the execution in a monitored environment.  
By transparency we mean: \emph{a program that executes all actions at the right times when running unmonitored will do so when running monitored}. Transparency and overhead are closely related in the timed scenario, since the overhead introduced by the monitor may interfere with the time in which the interactions are executed. We have tested the transparency by providing two different protocols - a protocol with resets and a protocol without resets. The former proves the usability of the monitor in a typical scenario, while the latter demonstrates its limitations. 
 
To set up the benchmark, we have fixed a Scribble timed protocol and manually created a correct implementation of the participants in that protocol using our timed Python API. We run the implementation in two scenarios using our monitoring framework, and with the monitors `turned off'. For each parameter configuration the protocol execution is repeated 30 times and the mean result is presented on a graph. Participants were run on the same machine (Intel(R) Core(TM) i7-2600 CPU $@$ 3.40GHz) to minimise the latency between the endpoints, hence test our framework in the more pessimistic scenario of small discrepancies between modelled delays and monitor overhead. All endpoint are connected via AMQP middleware broker and on average the latency between two endpoints is 0.04.
The full benchmark protocols, the applications and the raw data are available from the project page \cite{py_time}.

\paragraph{Scenario 1} We have initially considered a protocol with the same structure of the protocol in Figure~\ref{scribble_global} but with the constants in the clock constraints decreased by a scale of 100. We have changed the constraints to test transparency in a less optimistic scenario, with smaller difference between delays and monitor overhead (evidently transparency would also hold with larger differences, e.g., when using the constraints in Figure~\ref{scribble_global}). We used the implementation in Figure~\ref{master_role_python} with delays updated to match the protocol, as shown in Figure ~\ref{reset} (left). The outcome is presented in Figure \ref{reset} (right). The graph illustrates the time for completing a protocol for increasing number of recursive executions. 

This experiment shows that for the given protocol and implementation all executions are without constraint violation. Transparency is guaranteed (i.e., the overhead induced by the monitor does not affect the correctness of the program). Since resets prevent the monitor overhead to \emph{accumulate up to a non negligible overall delay} transparency is guaranteed even in case of a large number of iterations. The overhead introduced by the monitor is constant and due to the initial generation of the timed automaton from the textual Scribble timed local protocol (and just marginally to the checking of single interactions). 

\begin{figure}[t]
\vspace{-12mm}
\begin{minipage}{0.26\linewidth}
\begin{PYTHONLISTING}
global protocol ClientServer(
	role C, role S)
    {[x@C: x<c][x@S: x=c,]    
    ping(data) from C to S;
    {[x@C: x<2*c][x@S: x=2*c]    
    ping(data) from C to S;
    {[x@C: x<3*c][x@S: x=3*c]    
    ping(data) from C to S;
    {[x@C: x<3*c][x@S: x=4*c]    
    ping(data) from C to S;
    ...
    {[x@C: x<200*c][x@S: x=200*c]    
    ping(data) from C to S;
    
}
\end{PYTHONLISTING}
\end{minipage}
\begin{minipage}{0.22\linewidth}
\begin{PYTHONLISTING}
def server_proc(t):
    c = Conversation.
    	create(...)
    c.receive('S')
    while true:	
    	c.delay(t)
    	c.receive('W')
    	
    	
def server_proc(t, n):
    c = Conversation.
    	create(...)
    c.send('C')
    for i in range(0, n)	
    	c.delay(t)
    	c.send('C')
    	
\end{PYTHONLISTING}
\end{minipage}
\begin{minipage}{0.5\linewidth}
\includegraphics[scale=0.27]{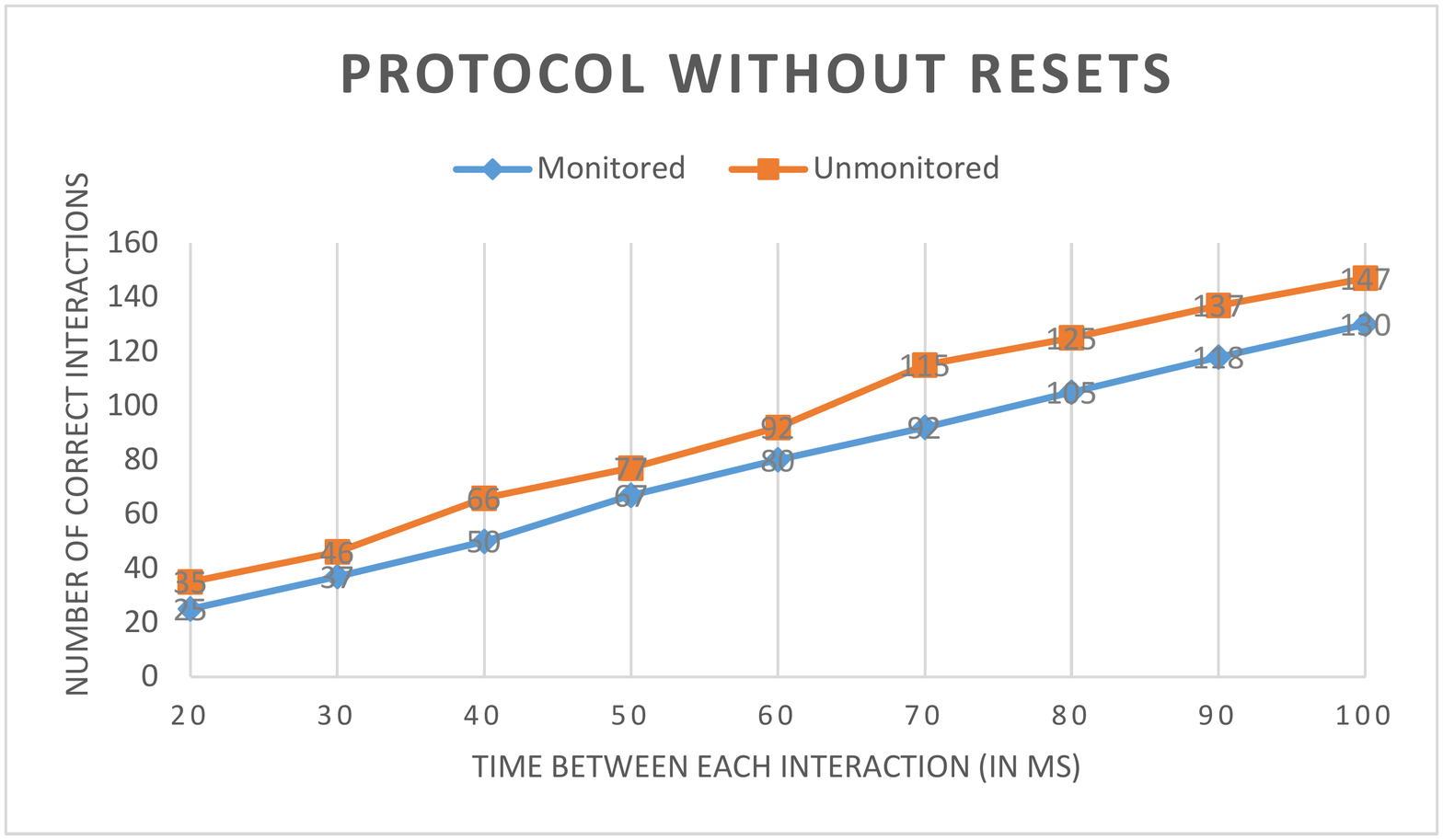}
\end{minipage}
\label{no_reset}
\vspace{-6mm}
\caption{Protocol with accumulating delays $c$ (left) and maximum number of correct interactions (right)}
\vspace{1mm}
\end{figure}

\paragraph{Scenario 2} Our second experiment was specifically targeted at checking how many interactions can generate a non-negligible accumulation of delays. 
We do this by removing resets. In case of no resets both the unmonitored and monitored programs are expected to start violating the constraints after certain number of executions. 
%
In Scenario 1 recursion allowed us to express repeated interactions by using resets. In order to observe a large number of repeated interactions \emph{without resets} we have created ad-hoc the sequential protocol in Figure~\ref{no_reset} (left) and implementation (middle). We have generated a protocol with 200 consecutive point to point interactions happening at increasing times (by $c$). We run the experiment for different values of $c$ (horizontal axis on the figure) and measure the maximum number of interactions (vertical axis on the figure) that can be executed before the program violates the time constraint.


The experiment confirmed that, the monitored application performs 90\% of the maximum number of possible interactions. This example comes to show the limitations of the timed monitoring framework. The practical scenarios we have encountered so far did not include long sequences of interactions, and repetitive operations are handled via recursions with resets at each cycle. 



\section{Related and Future Work}
\label{sec_conclusion}
The need for specifying and verifying the temporal requirements in a distributed systems is recognised. To this aim, different specification methods and verification tools have been developed, especially in the area of business process modelling (see \cite{survey} for a survey on verification of temporal properties). A work closely related to ours is \cite{WatahikiIH11}. It describes a framework for analysing choreographies between BPEL processes with time annotations. \cite{GuermoucheD12} extends BPML with time constraints and, via a  mapping from BPML to timed automata allows verification with the UPAAAL model checker. As a language for timed protocol specification, the main advantages of Scribble over alternatives such as BPEL, BPML and timed automata, is that it allows enforcement of global properties -- e.g., conformance of the interactions to global protocols -- while providing an in-built mechanism (projection) for decentralisation of the verification. 

Among the state-of-the-art runtime verification tools, a few support specification of temporal specifications \cite{Boer14,MOP,Larva}. \cite{MOP} presents a generic monitor that can be parametrised on the logic. \cite{Boer14} combines temporal properties and control flow specifications in a single formalism specified per object class. Our recovery mechanism resembles the aspect-oriented approaches used in those verifiers, but the combination of control flow checking and temporal properties in the same global specification is an unique characteristic of our work. Out tool checks statically the correctness of the specification itself in addition to the runtime checks for the program. Furthermore, via its formal basis, the framework 
allows to combine static verification and dynamic enforcement~\cite{FORTE13}.

\bibliographystyle{eptcs}
\bibliography{laurabib,session,time}

\end{document}